\documentclass[aps,twocolumn,prb,amsmath,amssymb,floatfix,superscriptaddress,reprint]{revtex4-2} 

\usepackage[utf8]{inputenc}
\usepackage{graphicx}
\usepackage{bm}
\usepackage{color}
\usepackage[colorlinks=true,citecolor=blue]{hyperref}
\usepackage{graphicx}

\usepackage{physics} 
\usepackage{comment}  
\makeatletter
\AtBeginDocument{\let\LS@rot\@undefined}
\makeatother

\DeclareMathOperator\Real{Re}
\DeclareMathOperator\Imag{Im}

\newcommand{\Deltastatic}{\Delta_\text{sc}}

\begin{document}

\title{Floquet engineering Higgs dynamics in time-periodic superconductors}
\author{Tobias Kuhn}
\affiliation{Department of Physics, University of Duisburg-Essen and CENIDE, D-47048 Duisburg, Germany}
\affiliation{Institute of Physics, University of Augsburg, D-86135 Augsburg, Germany}
\author{Björn Sothmann}
\affiliation{Department of Physics, University of Duisburg-Essen and CENIDE, D-47048 Duisburg, Germany}

\author{Jorge Cayao}
\affiliation{Department of Physics and Astronomy, Uppsala University, Box 516, S-751 20 Uppsala, Sweden}
\date{\today}

\begin{abstract}
 Higgs modes emerge in superconductors as collective excitations of the order-parameter amplitude when periodically driven by electromagnetic radiation.  In this work, we develop a Floquet approach to study Higgs modes in superconductors under time-periodic driving, where the dynamics of the order-parameter is captured by anomalous Floquet Green's functions.  We show that the Floquet description is particularly powerful as it allows one to exploit the time-periodic nature of the driving, thus considerably reducing the complexity of the time-dependent problem. Interestingly, the Floquet approach is also  enlightening  because it naturally offers a physical explanation for the renormalized steady-state order-parameter as a result of  photon processes between Floquet sidebands. We demonstrate the usefulness of Floquet engineering Higgs modes  in time-periodic $s$-wave superconductors.
\end{abstract}
\maketitle
\section{Introduction}
Superconductivity is a macroscopic quantum phenomenon that has attracted an enormous interest due to its relevance for future quantum technologies \cite{acin2018quantum,aguado2020perspective,aguado2020majorana,siddiqi2021engineering}. 
It emerges below a critical temperature due to the condensation of electron pairs also known as Cooper pairs, which are  characterized by a macroscopic complex wave function or order-parameter~\cite{Tinkham}. The superconducting order-parameter spontaneously breaks the continuous $U(1)$ gauge symmetry~\cite{landaux} and gives rise to collective excitations associated to its phase and amplitude \cite{PhysRevLett.4.380,RevModPhys.47.331,varma2002higgs,volovik2014higgs}. 
The phase  excitations, also known as Nambu-Goldstone modes,  are gapless but are shifted to the plasma frequency due to the Anderson-Higgs mechanism~\cite{PhysRev.130.439,higgs1964broken}. In contrast,  the amplitude excitations, also known as Higgs modes, are   gapped with an excitation energy equal to the superconducting energy gap~\cite{PhysRevLett.13.508,varma2002higgs}. In consequence, the Higgs modes represent the lowest-energy collective excitations of the order-parameter amplitude and, therefore, are central to the understanding of superconductivity~\cite{PhysRevB.84.174522,PhysRevB.87.054503,pashkin2014particle,pekker2015amplitude,Shimano_2020}.

The detection of Higgs modes in superconductors has been challenging   but 
promising   evidence has been recently reported~\cite{pekker2015amplitude,Shimano_2020}. The main difficulties  are due to {the} fact that Higgs modes are scalar excitations, which prevents their coupling to linear optical  probes \footnote{The absence of linear coupling between   Higgs modes and electromagnetic field  is consistent with the fact that Higgs modes do not have an electric charge and that their emergence is a second-order process \cite{pekker2015amplitude,Shimano_2020}.}, and also due to their low energies being of the order of the superconducting gap.  Despite {these} issues, it has been found that the presence of other competing orders, such as charge density waves, can make the Higgs modes detectable, for instance, by Raman spectroscopy~\cite{PhysRevLett.45.660,PhysRevB.23.3213,PhysRevB.89.060503,PhysRevB.90.224515,PhysRevB.97.094502}. Moreover, it has been predicted that intense light fields can excite Higgs modes even without other competing effects~\cite{volkov1974collisionless,PhysRevLett.93.160401,PhysRevB.72.220503,PhysRevLett.96.230404,PhysRevLett.96.097005,PhysRevLett.103.075301,PhysRevB.76.224522,PhysRevB.78.132505,PhysRevB.84.214513,PhysRevB.90.014515,PhysRevB.92.064508,PhysRevB.92.224517,PhysRevB.95.104507,PhysRevB.100.104515,PhysRevB.101.184519,PhysRevB.101.054502,PhysRevResearch.5.023011}, which has been recently reported  by using ultrafast THz pump-probe spectroscopy~\cite{PhysRevLett.111.057002,matsunaga2014light,sherman2015higgs,vaswani2021light,chu2020phase}. The advent of improved intense THz techniques~\cite{hebling2008generation,shimano2012intense,kampfrath2013resonant} therefore will facilitate the detection of Higgs modes in the future. Furthermore, it has been shown that Higgs modes permit to distinguish the symmetries of the superconductors~\cite{schwarz2020classification}, of pivotal relevance for understanding unconventional superconductivity and identifying possible quantum applications.

The importance of light fields to excite Higgs modes in superconductors has motivated the development of a time-dependent nonequilibrium framework, where the dynamics of the Higgs modes is described by a collective precession of Anderson pseudospins~\cite{PhysRev.112.1900,PhysRevB.92.064508}. In this case, superconductors under time-periodic driving signal the emergence of Higgs modes when the order-parameter amplitude oscillates with twice the driving frequency. At the same time, the amplitude of the oscillation exhibits a pronounced resonance when the driving frequency matches {the} superconducting gap energy. Even though the pseudospin Anderson description has been shown to be useful~\cite{Shimano_2020}, its application to  superconductors with more complicated structures is not straightforward. However, systems that are driven periodically in time can be conveniently  described with the help of Floquet theory~\cite{ASENS_1883_2_12__47_0,PhysRev.138.B979,PhysRevA.7.2203}, which is analogous to Bloch's theorem but formulated for the time domain and can, thus, reduce the complexity of the time-dependent problem. Despite this fact, however, it is still unknown how Floquet theory describes Higgs modes   in superconductors under time-periodic driving.

\begin{figure}
    \centering
    \includegraphics[width=\linewidth]{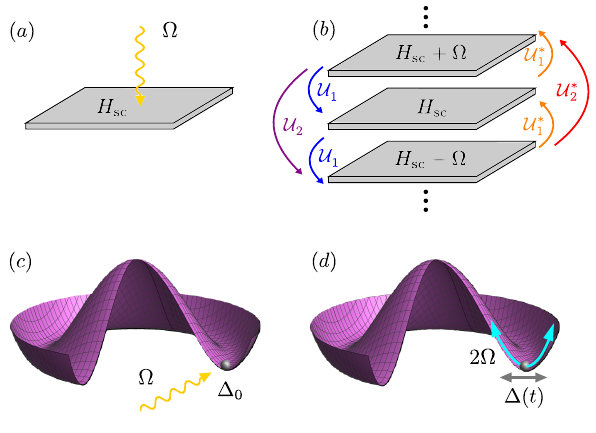}
    \caption{(a) A static superconductor (grey) with order-parameter $\Deltastatic$ described by the Hamiltonian $H_\text{sc}$ is periodically driven by a light field with frequency $\Omega$ depicted by the wiggle yellow arrow. (b)  The time dependent system  can be decomposed in terms of Floquet sidebands, labeled by sideband index $n$, where the system is described by replicas of the undriven Hamiltonian shifted in energy by  $n\Omega$ and coupled by $\mathcal{U}_{n}$ which depends on the applied light field. (c) Free energy of the static superconductor, where the continuous ground state symmetry breaking gives rise to a collective excitation,   known as Higgs mode, of the order-parameter amplitude $\Deltastatic$. Under the effect of a light field, the order-parameter becomes time-dependent and the    Higgs mode can be excited resonantly at energies $\Omega=\Deltastatic$.   (d) The time-dependent order-parameter $\Delta(t)$ becomes oscillatory with time and determines the Higgs dynamics, indicated by the cyan  double-headed arrow. Because Higgs modes only couple to light non-linearly, $\Delta(t)$ oscillates with a frequency of $2\Omega$.}
    \label{Fig0}
\end{figure}

In this work, we formulate a Floquet description of Higgs dynamics in time-periodic superconductors (see Fig.~\ref{Fig0}). In particular, we describe the dynamics of the superconducting order-parameter in terms of   anomalous Floquet Green's functions, which turns out to be a simple approach to explore the Higgs dynamics.  Since the Floquet description  maps a time-dependent system into a static problem by introducing Floquet sidebands, the approach developed here allows us to control the number of sidebands in the  Higgs dynamics.  To show the potential of the Floquet description, we  reproduce the resonant Higgs mode at driving energies equal to the superconducting gap in conventional $s$-wave superconductors and highlight its applicability to other superconductors. 
Interestingly, we find that the Floquet approach provides a natural and physical explanation for the renormalized order-parameter in the nonequilibrium steady-state regime, where the stationary order-parameter is renormalized by a nonequilibrium steady-state self-interaction (NESI) part that depends on transitions between Floquet sidebands via photon absorption and emission. The control and manipulation of the order-parameter by time-periodic drives paves the way for Floquet engineering Higgs dynamics in periodically driven superconductors. The remainder of this paper is organized as follows. We define the problem studied here in Sec.~\ref{section2}. In Sec.~\ref{section3}, we describe how pair amplitudes and the order-parameter in time-periodic superconductors are obtained within a Floquet description. In Sec.~\ref{section4}, we apply the Floquet method to study the order-parameter and Higgs dynamics  in conventional time-periodic superconductors.  Finally, in Sec.~\ref{section5}, we present our conclusions. To further support {the findings of this work}, in {Appendixes \ref{AppendixA} and \ref{AppendixB} we provide further details on the calculations of the Floquet Green's function in a finite Floquet space.}

\section{Defining the problem: Time-dependent order-parameter}
\label{section2}
We are interested in describing the dynamics of the order-parameter in superconductors under   time-periodic fields, which is expected to reveal the emergence of Higgs modes (see Fig.~\ref{Fig0}). In conventional spin-singlet $s$-wave superconductors, the time dependence of the order-parameter is then described by~\cite{PhysRev.108.1175},
\begin{equation}
     \label{eq:def-orderparameter}
    \hat\Delta(t)=\tilde{\lambda}\sum_{\pmb{k}}\langle c_{-\pmb{k},\downarrow}(t)c_{\pmb{k},\uparrow}(t)\rangle\,,
\end{equation}
where $\tilde{\lambda}$ is the constant attractive pairing interaction, $c_{\pmb{k},\sigma}$  annihilates an electronic state with spin $\sigma$, momentum $\pmb{k}$, at time $t$.  In the following, the $\hat{(.)}$ symbol  denotes time-dependent quantities unless otherwise specified. The sum on the right-hand side of Eq.~\eqref{eq:def-orderparameter} contains the anomalous average of two annihilation operators which is the fundamental characteristics of the superconducting state. 

The  anomalous averages seen above naturally appear when writing the system's Green's functions  in Nambu space  $\hat{\mathcal{G}}(\pmb{k};t,t')=-i\langle\mathcal{T}\Psi_{\pmb{k}}(t)\Psi_{\pmb{k}}^\dagger(t')\rangle$,   where $\Psi_{\pmb{k}}=(c_{\pmb{k}\uparrow},c_{-\pmb{k}\downarrow}^\dagger)^T$ is the Nambu spinor and $\mathcal{T}$ the time-ordering operator \cite{mahan2013many,zagoskin1998quantum}. Then, $\hat{\mathcal{G}}(\pmb{k};t,t')$ is given by  
\begin{equation}
\label{eq:green}
  \hat{\mathcal{G}}(\pmb{k};t,t')=\begin{pmatrix}
    \hat G(\pmb{k};t,t')&\hat F(\pmb{k};t,t')\\
    \hat F^\dagger(\pmb{k};t,t')&\hat G^\dagger(\pmb{k};t,t')
    \end{pmatrix}\,,
 \end{equation}
where $\hat G(\pmb{k};t,t')=  -i\langle\mathcal{T}c_{\pmb{k}\uparrow}(t)c_{\pmb{k}\uparrow}^\dagger(t')\rangle$ is the normal component and $\hat F(\pmb{k};t,t')=-i\langle\mathcal{T}c_{\pmb{k}\uparrow}(t)c_{-\pmb{k}\downarrow}(t')\rangle$ the anomalous pair correlation of the  Green's function~\cite{mahan2013many,zagoskin1998quantum}. Now, by a direct comparison between Eqs.~\eqref{eq:green} and~\eqref{eq:def-orderparameter}, the time-dependent order-parameter $\Delta(t)$  can immediately be defined in terms of the pair correlations as,
 \begin{equation}
     \label{eq:op floq dyn}
    \hat\Delta(t)=i\tilde{\lambda}\sum_{\pmb{k}} \hat F(\pmb{k};t,t)\,,
\end{equation}
where $\hat F(\pmb{k};t,t)$ is the anomalous  component of the Nambu Green's function in Eq.~\eqref{eq:green} evaluated at $t'=t$.  As discussed at the beginning of this section, we are interested in the dynamics of the order-parameter amplitude $\hat \Delta(t)$ and in its Higgs mode  when time-periodic drivings are applied.  Eq.\,(\ref{eq:op floq dyn}) shows that, to address the dynamics of the order-parameter, it is necessary to describe and understand the time dependence of the pair amplitudes $\hat F(\pmb{k};t,t)$ under time-periodic driving which is the problem we aim to address in this work. We note that although the above   discussion has been formulated for spin-singlet superconductors, the relationship between pair amplitudes and order-parameter also holds for spin-triplet superconductors; the only difference is that the pair amplitudes then become matrices in spin space, thus enabling the emergence of spin-triplet components~\cite{RevModPhys.63.239,cayao2019odd}. Below, we show how  the pair amplitudes and order-parameter can be obtained by exploiting their time-periodicity within Floquet theory.

\section{Floquet   pair amplitudes and order-parameter dynamics} 
\label{section3}
In this section we employ Floquet theory to describe the pair amplitudes of time-periodic superconductors,  which correspond to the anomalous part of the Nambu Green's function $ \hat{\mathcal{G}}(\pmb{k};t,t')$ given by Eq.~\eqref{eq:green}.  For this purpose,  we first aim at finding $\hat{\mathcal{G}}(\pmb{k};t,t')$,  which is obtained by solving the equation of motion $[i\partial_{t} - \hat H_{\boldsymbol{k}}(t)]\hat{\mathcal{G}}(\boldsymbol{k};t,t')=\delta(t-t')$, where $\hat H_{\boldsymbol{k}}(t)$ is the Hamiltonian of a  time-periodic superconductor in Nambu space.

\subsection{Floquet Green's function and Floquet pair amplitudes} 
\label{subsection3a}
We consider time-periodic superconductors which emerge as a result of exposing a static superconductor described by a  Nambu Hamiltonian $H_{\rm sc}(\pmb{k})$ to a time-periodic drive $\hat{\mathbf E}(t)$ with period $T=2\pi/\Omega$ and frequency $\Omega$, see Fig.~\ref{Fig0}.  For now, we assume that $H_{\rm sc}(\pmb{k})$ describes a generic superconductor and its explicit form will be given later. The effect of the time-dependent drive $\hat{\mathbf E}(t)$ is introduced by a minimal coupling substitution   $\pmb{k}\rightarrow \pmb{k} + e\hat{\mathbf A}(t)$, where $\hat{\mathbf A}(t)$ is the vector potential $\hat{\mathbf E}(t)=-\partial_{t}\hat{\mathbf A}(t)$ and $e>0$ is the elementary electron charge. The total time-dependent Hamiltonian can {then} be written as $\hat H_{\pmb{k}}(t)=H_{\rm sc}(\pmb{k})+\hat V_{\pmb{k}}(t)$ where   $H_{\rm sc}(\pmb{k})$ describes the undriven superconductor, while  $\hat V_{\pmb{k}}(t)$ entirely depends on the drive $\hat{\mathbf{E}}(t)$, and its {explicit} form will be discussed later. Then, the total Hamiltonian $\hat H_{\pmb{k}}(t)$ acquires the time dependence of   $\hat{\mathbf{E}}(t)$ and becomes periodic in time, namely,  $\hat H(t)=\hat H(t+T)$. For this type of time-periodic Hamiltonians, the Floquet theorem permits us to write the solutions of the Schr\"{o}dinger equation in terms of  harmonics of the driving frequency $\Omega$ referred to as Floquet modes~\cite{ASENS_1883_2_12__47_0,PhysRev.138.B979,PhysRevA.7.2203},  see also Ref.\,\cite{PhysRevLett.110.200403}. 

In the Floquet picture, the time-periodic Hamiltonian can be decomposed in Floquet modes as $\hat H(t)=\sum_{m}H_m{\rm e}^{-im\Omega t}$, while the Green's function  $\hat{\mathcal{G}}(t,t')=\hat{\mathcal{G}}(t+T,t'+T)$ can be written as~\cite{RevModPhys.86.779}
 \begin{equation}
\label{floquetF}
 \hat{\mathcal{G}}(\pmb{k};t,t')=
 \sum_{m,n}\int \frac{d\omega}{2\pi} {\rm e}^{-i(\omega +m\Omega)t+i(\omega +n\Omega)t' } \mathcal{G}_{m,n}(\boldsymbol{k},\omega)\,,
\end{equation}
 where the coefficients $\mathcal{G}_{n,m}$ represent the Floquet Green's function amplitudes, labeled by the Floquet indices $n,m\in\mathbb{Z}$, and $\omega\in[-\Omega/2,\Omega/2]$. We can write  the equation of motion for $\hat{\mathcal{G}}$ in Floquet space as~\cite{RevModPhys.86.779}
\begin{equation}
\label{eq:eom}
\begin{split}
\sum_{m'}\big[\omega\delta_{m,m'}-\mathcal H_{m,m'}\big]\mathcal{G}_{m',n}({\boldsymbol{k}},\omega)=\delta_{m,n}\,,
\end{split}
\end{equation}
where
\begin{equation}
\label{Floquetcouplings}
\begin{split}
\mathcal H_{m,n}&=\left(H_\text{sc}-n\Omega\right)\delta_{m,n}+\mathcal U_{m,n}\,, \\
\mathcal U_{m,n}&=\frac{1}{T}\int_{0}^{T}dt{\rm e}^{i(m-n)\Omega t} \hat V_{\boldsymbol{k}}(t)\,,
 \end{split}
 \end{equation}
and  $m$ and $n$ represent Floquet indices. In deriving the equation of motion, we used $\delta_{m,n}=(1/T)\int_{0}^{T}dt{\rm e}^{i(m-n)\Omega t}$ and omitted the momentum label in the Floquet Hamiltonian harmonics $\mathcal H_{m,n}$ for brevity.
Thus, we have obtained an equation of motion in terms of Floquet modes $\mathcal H_{m,n}$ and $\mathcal{G}_{m,n}$ which does not involve any time dependence as a result of employing the Floquet decomposition. The mathematical structure of the equation of motion can be visualized as shown in Fig.~\ref{Fig0} (b). The diagonal elements $H_\text{sc}+n\Omega$ describe replicas of the original Hamiltonian $H_{\rm sc}$ shifted by integer multiples of the driving frequency $\Omega$. The off-diagonal components $\mathcal{U}_{m,n}$ couple the Floquet bands, are determined by the driving, and involve the emission ($n>m$) or absorption ($n<m$) of $|n-m|$ photons. We also note that, while the sum over Floquet harmonics in Eq.~\eqref{eq:eom} runs, in principle, to infinity, it can be safely truncated due to the focus on a finite range of frequencies $\omega$ and still  approximate well the exact result~\cite{RevModPhys.86.779,rudner2020band,rudner2020floquet}. 
The determination of the Floquet Green's function components $\mathcal{G}_{m,n}$ via the equation of motion~\eqref{eq:eom} then involves a finite matrix inversion. The Floquet components of the Green's functioncan then be used to calculate the time-dependent Green's function $\hat{\mathcal{G}}(\pmb{k};t,t')$ by means of Eq.~\eqref{floquetF}. 

Having found the time-dependent Green's function using Floquet modes, we are now in position to discuss  the calculation of the Floquet pair amplitudes which will allow us to obtain the order-parameter dynamics. The Nambu structure of the static Hamiltonian $H_{\rm sc}$ is inherited by the Fourier harmonics $\mathcal H_{m,n}$ and $\mathcal{G}_{m,n}$. Therefore, the off-diagonal  components of the Floquet Green's function $\mathcal{G}_{m,n}$  in Nambu space give the Floquet pair amplitudes which we denote as $F_{m,n}$. These Floquet pair amplitudes were shown to naturally appear in time-periodic superconductors~\cite{PhysRevB.103.104505}  where they provide a physical interpretation of different emergent superconducting pairs between Floquet  bands due to emission and absorption of photons. 
  
\subsection{Order-parameter dynamics from Floquet pair amplitudes in the time domain}
\label{subsection3d}
Using the Floquet representation of the anomalous Green's function, we can write the time-dependent order-parameter as
\begin{equation}
    \label{eq:floquet_orderparameter}
    \hat\Delta(t)=i\tilde{\lambda}\sum_{\pmb{k},m,n}\int_{-\Omega/2}^{\Omega/2}\mathrm{d}\omega\;  F_{m,n}(\pmb{k},\omega)\,e^{-i(m-n)\Omega t}.
\end{equation}
As the order-parameter depends only on the pair amplitude evaluated at equal times, the order-parameter oscillates with integer multiples of the driving frequency $\Omega$ only and does not depend on $\omega$. In particular, when the number of Floquet bands is cut off when determining the Floquet Green's function $F_{m,n}(\pmb{k},\omega)$, the maximal oscillation frequency of the order-parameter is given by number of Floquet bands multiplied with the driving frequency.
The time evolution of the order-parameter can be decomposed into its Fourier components as
\begin{equation}
    \label{eq:floquet_orderparameter2}
    \hat \Delta(t)=\sum_{l}\Delta_{l}(\Omega) e^{il\Omega t},
\end{equation}
where
\begin{equation}
\label{eq:floquet_orderparameter3}
\Delta_{l}(\Omega)=i\tilde{\lambda}\sum_{\pmb{k},m}\int_{-\Omega/2}^{\Omega/2}\mathrm{d}\omega\;  F_{m+l,m}(\pmb{k},\omega)\,.
\end{equation}
As can be shown by a straightforward calculation (cf. Appendix~\ref{sec:dDl proof}) one has $\Delta_l(\Omega)=\Delta_{-l}^\ast(\Omega)$ which ensures that the order-parameter is real. According to Eq.~\eqref{eq:floquet_orderparameter2}, the order-parameter oscillates around its average value $\Deltastatic(\Omega)$ with amplitudes $\Delta_l(\Omega)$ and frequency $l\Omega$. The behavior seen here for $\hat \Delta(t)$ is analogous to what is obtained in the Anderson's pseudospin picture where the dynamics of the order-parameter is also dictated by deviations from the static regime. Therefore, the amplitudes $\Delta_{l}$  in Eq.~\eqref{eq:floquet_orderparameter3} describe the dynamics of the Higgs mode. We remark that in the static regime without any external driving, the left-hand side of Eq.~\eqref{eq:floquet_orderparameter} must yield the order-parameter in the static regime which we denote by $\Deltastatic$. However, in the presence of external driving, the average value of the order-parameter can deviate from its static value due to a nonequilibrium renormalization caused by the coupling to other Floquet bands.  This nonequilibrium self-interaction (NESI) is generally non-zero and shifts the order-parameter in a time-independent fashion. We can characterize this NESI state by
\begin{equation}
\Delta_\text{NESI}=\Delta_0(\Omega)-\Deltastatic\,,
\label{eq:NESI}    
\end{equation}
which involves the contributions of all relevant Floquet sidebands. We note that this effect was already pointed out when analyzing the Higgs dynamics within the Anderson's pseudospin description but its interpretation was not further discussed~\cite{PhysRevB.92.064508}. The Floquet bands employed here, however, naturally reveal that such self-interaction emerges as a result of photon-assisted pair correlations between Floquet bands with equal Floquet indices.

\subsection{Floquet pair amplitudes for large driving frequencies}  
\label{subsection3e}
In order to obtain the Floquet pair amplitudes  $F_{n,m}$ which characterize the order-parameter dynamics, one has solved Eq.~\eqref{eq:eom}. In principle, this involves the inversion of an infinite-dimensional matrix. However, as we have pointed out above, one is usually interested in a finite frequency range $\omega$ only such that it is possible to neglect higher Floquet bands. Various previous works have demonstrated that   one can obtain good results for a variety of time-dependent problems when taking into account only $n=0,\pm 1,\pm 2$~\cite{rudner2020floquet,rudner2020band,RevModPhys.86.779,PhysRevB.103.104505}. While the restriction to a finite number of Floquet bands already simplifies the matrix inversion, an additional simplification can be achieved when the driving frequency is much larger than the coupling between Floquet bands, $\mathcal{U}_{m,n}/\Omega\ll 1$. In this limit, one can perform a systematic perturbation theory in the Floquet-band coupling which allows one not only to calculate the Flqouet pair amplitudes but also provides an intuitive way to visualize the functional dependencies of the pair amplitudes. To this end, we exploit the Dyson equation for each Floquet Green’s function component which reads as $\mathcal{G}=g+gV\mathcal{G}$,  where $\mathcal{G}$ is the dressed Green’s function, $g$ represents the undressed propagator in the respective sideband and $V$ denotes the coupling between sidebands. Up to second order in $V$, the previous equation can be written as $\mathcal{G}\approx g+gVg+gVgVg$. Then, by projecting this second order approximation onto Floquet bands, we get
\begin{equation}
 \label{eq:dyson_GF}
 \mathcal{G}_{m,n}\approx\bra{m}g\ket{n}+\bra{m}gVg\ket{n}+\bra{m}gVgVg\ket{n}\,,
\end{equation}
where $\ket{n}$ and $\ket{m}$ denote Floquet bands, $g_{m,n}=\bra{m}g\ket{n}$ is the projection of the intraband propagator onto Floquet bands which is finite only for $m=n$, and $V_{m,n}=\bra{m}V\ket{n}$ describes the coupling between sidebands. The value of the coupling depends on the structure of the applied drive $\hat{\mathbf{E}}(t)$ (see also previous two subsections). We remark that all the elements of Eq.~\eqref{eq:dyson_GF} are matrices in Nambu space, such that the Floquet pair amplitudes $F_{m,n}$ correspond to the off-diagonal components of $\mathcal{G}_{m,n}$. Thus, using a perturbation approach, it is possible to obtain further understanding of the Floquet pair amplitudes, especially about their functional dependencies. While Eq.~\eqref{eq:dyson_GF} has been formulated up to second order in perturbation theory, it can readily be extended to include higher orders and an arbitrary number of Floquet sidebands.

\section{Floquet Higgs dynamics in conventional time-periodic superconductors} 
\label{section4}
In the following, we illustrate our general Floquet theory of Higgs dynamics with the example of a conventional spin-singlet $s$-wave superconductor which is subject to a time-periodic driving by an external electric field. The static superconductor is  modeled by
\begin{equation}
\label{H0SC}
    H_{\rm sc}(\pmb{k})=\xi_{\pmb{k}}\tau_z+\Deltastatic\tau_x\,,
\end{equation}
where   $\xi_{\pmb{k}}={\pmb{k}^2}/{2m}-\mu$ is the kinetic energy with chemical potential $\mu$,  $\pmb{k}=(k_{x},k_{y},k_{z})$ denotes the momentum, $\tau_{i}$ represents the $i$-th Pauli matrix in Nambu space, and $\Psi_{\pmb{k}}=(c_{\pmb{k}\uparrow},c_{-\pmb{k}\downarrow}^\dagger)^T$. Here, $\Deltastatic$ represents the spin-singlet $s$-wave order-parameter, chosen to be real without loss of generality. For the time-periodic driving, we consider linearly polarized light with a vector potential given by $\hat{\mathbf{A}}(t)=A_{0}({\rm sin(\Omega t),0,0})$, which has a period $T=2\pi/\Omega$. Then, the effect of the driving is incorporated by a minimal coupling substitution $\pmb{k}\rightarrow\pmb{k}+e\hat{\mathbf A}(t)$ with $e>0$, which leads to the a time-dependent Hamiltonian given by $\hat H_{k_x}(t)=H_{\rm sc}+\hat V_{k_x}(t)$, where $H_{\rm sc}$ is given by Eq.~\eqref{H0SC} and 
\begin{equation}
\hat V_{k_x}(t)=\frac{e k_{x} A_{0}}{m}{\rm sin}(\Omega t)\tau_{0}-\frac{e^{2}  A_{0}^{2}}{4m}{\rm cos}(2\Omega t)\tau_{z}\,,
\label{eq:drive-potential}
\end{equation}
and we have renormalized the chemical potential as $\mu\to\mu-e^2A_0^2/4m$. We see that the total system Hamiltonian $\hat H(t)$ is indeed time periodic $\hat H_{k_x}(t)=\hat H_{k_x}(t+T)$. We are thus in position to apply the Floquet approach developed in the previous section to describe the order-parameter and Higgs dynamics given by Eqs.~\eqref{eq:floquet_orderparameter}, \eqref{eq:floquet_orderparameter2}, \eqref{eq:floquet_orderparameter3}. To this end, we first calculate the Floquet pair amplitudes $F_{m,n}$ by solving the equation of motion~\eqref{eq:eom}. For the chosen driving field, the coupling between the Floquet bands is only finite for nearest and next-nearest neighbor sidebands,
\begin{equation}
\begin{split}
\mathcal{U}_{m,n}&=\mathcal{U}_{1}\delta_{m+1,n}+\mathcal{U}_{2}\delta_{m+2,n}\\
&+\mathcal{U}_{1}^{*}\delta_{m-1,n}+\mathcal{U}_{2}^{*}\delta_{m-2,n}\,,
\end{split}
\label{eq:hopping-amplitudes}
\end{equation}
where $\mathcal{U}_{1}= U_1\tau_{0}$, $\mathcal{U}_{2}= U_2\tau_{z}$,  $U_{1}={i ek_{x}A_{0}}/(2m)$, and $U_{2}=-{e^{2}A^{2}_{0}}/(8m)$ depend on the driving amplitude $A_{0}$. We use Eq.~\eqref{eq:dyson_GF} to find the components  of the Floquet Green's functions $\mathcal{G}_{n,m}$ within second-order perturbation theory in the coupling between sidebands which allows us to obtain compact expressions for the pair amplitudes.  As we have pointed out above, the perturbation theory can be applied if $\mathcal{U}_{n}/\Omega\ll1$, i.e. for weak driving amplitudes. Within this approximation, we obtain the Floquet pair amplitudes $F_{m,n}(\pmb{k},\omega)$ as
\begin{widetext}
    \begin{equation}
    \label{FloquetpairF}
\begin{split}
F_{0,0}&=\frac{\Deltastatic}{D_0}+\frac{\Deltastatic |U_{1}|^2}{D_0^2D_{-1}D_{1}}[2(\Omega^2+D_0)(4\omega^2-D_0)-8(\omega\Omega)^2]+\frac{\Deltastatic |U_{2}|^2}{D_0^2D_{-2}D_{2}}[2(4\Omega^2+D_0)(4\xi_{\pmb{k}}^2+D_0)-8(2\omega\Omega)^2]\,,\\
F_{1,1}&=\frac{\Deltastatic}{D_1}+\frac{\Deltastatic |U_{1}|^2}{D_0D_1^2}[4\omega^2-D_0+\Omega^2+4\omega\Omega]+\frac{\Deltastatic |U_{1}|^2}{D_2D_1^2}[4(\omega+\Omega)^2+\Omega^2-D_0]+\frac{\Deltastatic |U_{2}|^2}{D_{-1}D_1^2}[D_{-1}+4\xi_{\pmb{k}}^2-4\Omega^2]\,,\\
F_{-1,-1}&=\frac{\Deltastatic}{D_{-1}}+\frac{\Deltastatic |U_{1}|^2}{D_0D_{-1}^2}[4\omega^2-D_0+\Omega^2-4\omega\Omega]+\frac{\Deltastatic |U_{1}|^2}{D_{-2}D_{-1}^2}[4(\omega-\Omega)^2+\Omega^2-D_0]+\frac{\Deltastatic |U_{2}|^2}{D_{1}D_{-1}^2}[D_{1}+4\xi_{\pmb{k}}^2-4\Omega^2]\,,\\
F_{2,2}&=\frac{\Deltastatic}{D_{2}}+\frac{\Deltastatic |U_{1}|^2}{D_1D_{2}^2}[4(\omega+\Omega)^2+2\omega\Omega+4\Omega^2-D_0]+\frac{\Deltastatic |U_{2}|^2}{D_{0}D_{2}^2}[D_{0}+4\xi_{\pmb{k}}^2-4\Omega^2]\,,\\
F_{-2,-2}&=\frac{\Deltastatic}{D_{-2}}+\frac{\Deltastatic |U_{1}|^2}{D_{-1}D_{-2}^2}[4(\omega-\Omega)^2-2\omega\Omega+4\Omega^2-D_0]+\frac{\Deltastatic |U_{2}|^2}{D_{0}D_{-2}^2}[D_{0}+4\xi_{\pmb{k}}^2-4\Omega^2]\,,\\
F_{1,-1}&=\frac{2\Deltastatic U_2^\ast}{D_1D_{-1}}[\xi_{\pmb{k}}+\Omega]+\frac{\Deltastatic ({U}_{1}^{*})^2}{D_{1}D_{0}D_{-1}}[D_0+\Omega^2-4\omega^2]\,,\\
F_{-1,1}&=\frac{2\Deltastatic U_2}{D_{-1}D_{1}}[\xi_{\pmb{k}}-\Omega]+\frac{\Deltastatic {U}_{1}^2}{D_{1}D_{0}D_{-1}}[D_0+\Omega^2-4\omega^2]\,,\\
F_{0,2}&=\frac{2\Deltastatic U_2}{D_{0}D_{2}}[\xi_{\pmb{k}}-\Omega]+\frac{\Deltastatic {U}_{1}^2}{D_{0}D_{1}D_{2}}[D_0-2\Omega^2-4\omega^2-6\omega\Omega]\,,\\
F_{-2,0}&=\frac{2\Deltastatic U_2}{D_{-2}D_{0}}[\xi_{\pmb{k}}-\Omega]+\frac{\Deltastatic {U}_{1}^2}{D_{-2}D_{-1}D_{0}}[D_0-2\Omega^2-4\omega^2+6\omega\Omega]\,,\\
F_{2,0}&=\frac{2\Deltastatic U_2^\ast}{D_2D_{0}}[\xi_{\pmb {k}}+\Omega]+\frac{\Deltastatic ({U}_{1}^{*})^2}{D_{2}D_{1}D_{0}}[D_0-2\Omega^2-4\omega^2-6\omega\Omega]\,,\\
F_{0,-2}&=\frac{2\Deltastatic U_2^\ast}{D_0D_{-2}}[\xi_{\pmb{k}}+\Omega]+\frac{\Deltastatic ({U}_{1}^*)^{2}}{D_{0}D_{-1}D_{-2}}[D_0-2\Omega^2-4\omega^2+6\omega\Omega]\,,
\end{split}
\end{equation}
\end{widetext}
where  $D_n\equiv(\omega+n\Omega)^2-\xi_{\pmb{k}}^2-\Deltastatic^2=D_0+2n\omega\Omega+(n\Omega)^2$ and  for simplicity we have dropped the arguments $(\omega,\pmb{k})$ in the Floquet pair amplitudes.  Here $\xi_{\pmb{k}}$ is the kinetic energy introduced in Eq.\,(\ref{H0SC}). {Despite the apparent complexity in the expressions above, these pair amplitudes exhibit a natural physical interpretation. In fact, the pair amplitudes $F_{n,n}$ represent intra-sideband  pair correlations and are determined by the bare pair amplitudes (first term on the right hand side) and corrections  due to transitions between sidebands, which are proportional to $|U_{1,2}|^{2}$ and are assisted by  photon processes with an equal number of emitted and absorbed photons. In contrast, the pair amplitudes $F_{n,m}$ with $n\neq m$ represent inter-sideband pair correlations, with transitions determined by $U_{2}$ and $(U_{1})^{2}$, or by its conjugate, thus involving either absorption or emission of photons. We note that} these pair amplitudes are consistent with those found in Ref.~\cite{cayao2019odd}, but acquire additional components because the considered drive is linearly polarized in contrast to the circularly polarized drive considered in Ref.~\cite{cayao2019odd}. We remark that it is straightforward to obtain similar expressions for the pair amplitudes when taking into account additional Floquet bands (see Appendix~\ref{sec:propagatorelements}).
However,  to understand the Higgs dynamics and NESI state, it is sufficient to take into account Floquet pair amplitudes $\mathcal F_{m,n}$ with $|m-n|=0,\pm2$. 

\begin{figure}
    \centering
    \includegraphics[width=\linewidth]{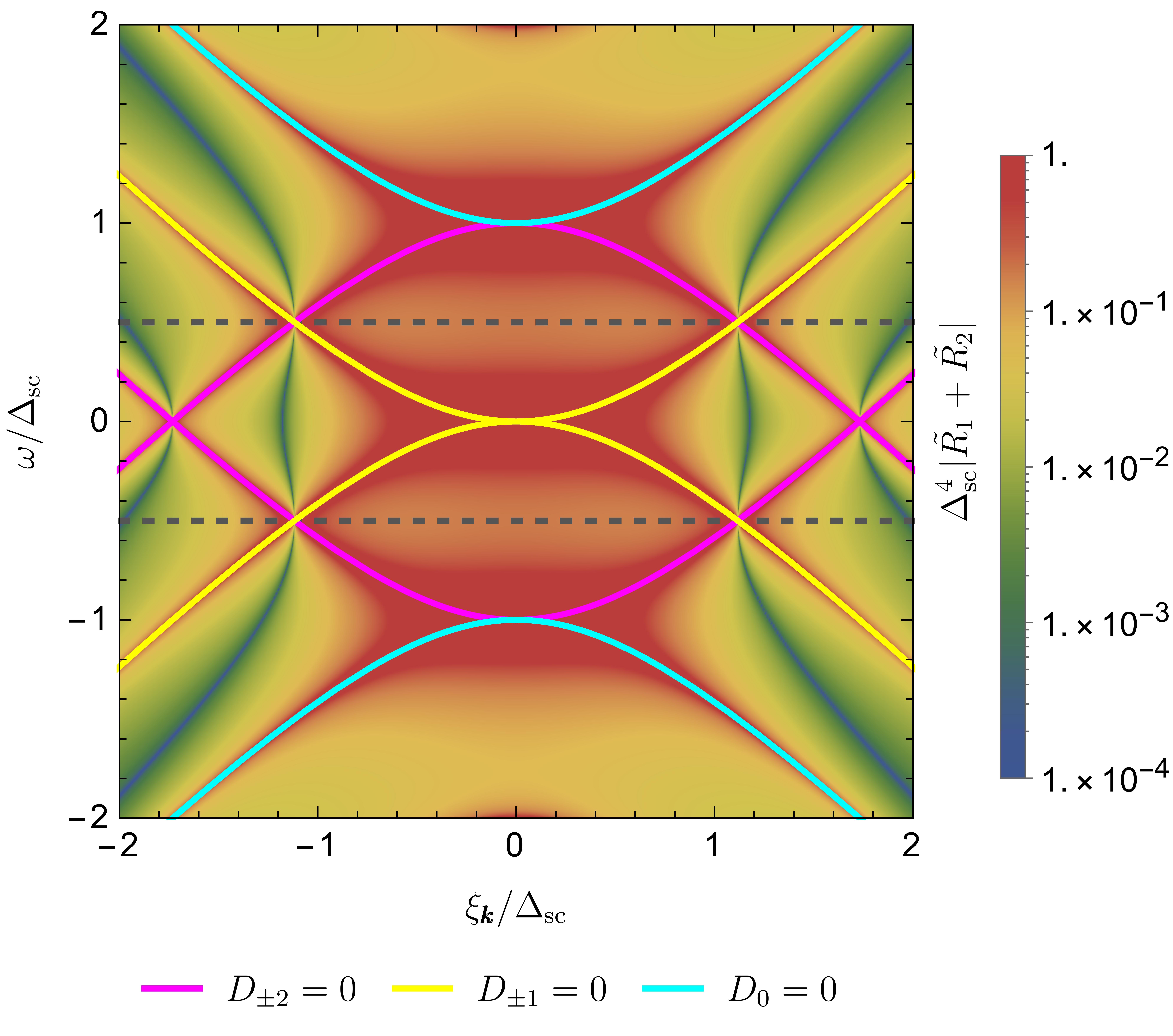}
    \caption{Floquet pair amplitude of sideband difference $|l|=2$ characterized by the integrand $\tilde R_1(\Omega)+\tilde R_2(\Omega)$ of Eq.~\eqref{RR} as a function of   frequency $\omega$ and  dispersion $\xi_{\pmb{k}}$. Moreover, the Floquet bands are plotted after solving $D_n=0$ defined below Eq.\eqref{FloquetpairF} for $n=0,\pm1,\pm2$, shown by cyan, yellow, and magenta curves, respectively.   Note that $\tilde{R}_1$ and $\tilde{R}_2$ contain the contributions of  $n=\pm1$ and  $n=0,\pm2$, respectively. Dashed gray lines indicate the integration boundaries of Eq.~\eqref{RR}. Here,  driving at frequencies of the static order-parameter $\Omega=\Deltastatic$ exhibits a dominant central singularity.}
    \label{fig:FloquetBands}
\end{figure}

Before analyzing the Floquet Green's functions and the associated time-dependent order-parameter in detail, in Fig.~\ref{fig:FloquetBands} we plot the Floquet bands by solving $D_n=0$ for $\omega$ and $n=0,\pm1,\pm2$; the bands are depicted by cyan, yellow, and magenta curves, respectively. It is worth noting that the sidebands $n=\pm1$ (purple) meet at $\omega=0$ to form a singularity $\sim1/(\Omega^2-\Deltastatic^2)^2$ which is the dominant singularity in the frequency and energy range of interest. Below we will see that it is this singularity which gives rise to the resonant behavior of the order-parameter amplitude at $\Omega=\Deltastatic$ which then results in the resonant Higgs mode.

\subsection{Dynamical contribution to Higgs modes}
By using the Floquet pair amplitudes calculated in Eq.\,\eqref{FloquetpairF}, we now calculate the order-parameter dynamics using Eq.~\eqref{eq:floquet_orderparameter2} to describe the Higgs dynamics. Due to momentum symmetry, all contributions odd in $\pmb{k}$ vanish and lead to the well known fact that Higgs modes do not couple linearly to light in the framework of Floquet engineering. Additionally, all odd $\xi_{\pmb{k}}$ as well as odd $\omega$ pair amplitudes do not produce an observable signal either because of symmetrical integration bounds. The remaining contribution is dominated by next-nearest neighbor couplings between Floquet pair amplitudes $\mathcal F_{m,n}$ with $|m-n|=\pm2$. Thus, the time-dependent order-parameter in Eq.~\eqref{eq:floquet_orderparameter2}  oscillates with $2\Omega$ and has an amplitude
    \begin{equation}
    \label{deltaD2}
        \begin{split}
            \Delta_{2}(\Omega)=&
            i\lambda\int^{\varepsilon_C}_{-\varepsilon_C}\mathrm{d}\xi_{\pmb{k}}\int_{-\Omega/2}^{\Omega/2}\mathrm{d}\omega  (F_{-2,0}+F_{-1,1}+F_{0,2})\\
            =&-i\lambda\int\displaylimits^{\varepsilon_C}_{-\varepsilon_C}\mathrm{d}\xi_{\pmb{k}}\int\displaylimits_{-\Omega/2}^{\Omega/2}\mathrm{d}\omega  
            \left[\frac{2\Deltastatic U_2\Omega}{D_{-1}D_{1}}\right.\\&\hspace{26mm}+\left.\frac{4\Deltastatic U_2\Omega(D_0+4\Omega^2)}{D_{-2}D_0D_2} \right]\,,
        \end{split}
    \end{equation}
where in the second equality we have used the expressions for the pair amplitudes $F_{-2,0}$, $F_{-1,1}$, and $F_{0,2}$ given in Eqs.~\eqref{FloquetpairF}. To obtain Eq.\,\eqref{deltaD2}, we replaced the sum over momenta by an energy integration assuming a constant density of states $D_F$ near the Fermi energy and introduced $\lambda=\tilde{\lambda}D_F$.
Furthermore, we introduced the Debye energy $\varepsilon_C$ as an appropriate cutoff for the energy integration. Using $\Delta_2(\Omega)=\Delta^*_{-2}(\Omega)$, we can write the time-dependent order-parameter as
    \begin{equation}
           \label{eq:HiggsDyn_ConvSC}
    \begin{split}
         \hat \Delta(t)-\Delta_{0}(\Omega)&=\sum_{l\neq0}\Delta_l(\Omega)e^{il\Omega t}\\
            &=4\Deltastatic U_2\lambda R(\Omega)\sin{(2\Omega t)}\,,\\
               \end{split}\end{equation}
where $\Delta_{0}(\Omega)$ contains the nonequilibrium renormalization of the static order-parameter discussed above Eq.~\eqref{eq:NESI}, while      
\begin{equation}
\label{RR}
R(\Omega)=\sum_{i=1,2}R_i(\Omega)\,,
\end{equation}
where
\begin{equation}
R_{i}=\Omega\int\displaylimits^{\varepsilon_C}_{-\varepsilon_C}\mathrm{d}\xi_{\pmb{k}}\int\displaylimits_{-\Omega/2}^{\Omega/2}\mathrm{d}\omega\;  \tilde{R}_{i}(\Omega)\,,
\end{equation}
and
\begin{equation}
    \label{tildeR}
    \begin{split}      
        \tilde{R}_{1}(\Omega)&=\frac{1}{D_{-1}D_{1}}\,,\quad\tilde{R}_{2}(\Omega)=\frac{2[D_0+4\Omega^2]}{D_{-2}D_0D_2}\,.    
    \end{split}
\end{equation}
It is evident that the quantities $\tilde{R}_{1(2)}$ contain contributions from sidebands $n=\pm1$ ($n=0,\pm2$) (see also  Fig.\,\ref{fig:FloquetBands}).
Equation~\eqref{eq:HiggsDyn_ConvSC} describes the dynamics of the order-parameter in the time domain. As such, it describes the dynamics of the Higgs mode. By a direct inspection of Eq.~\eqref{eq:HiggsDyn_ConvSC} we observe that the order-parameter oscillates with twice the driving frequency, $2\Omega$. The amplitude of these oscillations is determined by the static order-parameter $\Deltastatic$, the effective electron-electron interaction $\lambda$ and the coupling between Floquet bands $U_{2}$ which involves the strength of the driving field. In addition, it depends on the function $R(\Omega)$ given by Eq.~\eqref{RR} which dictates the nontrivial dependence of the order-parameter dynamics on the driving frequency and, thus, contains the key information about the Higgs dynamics.  We note that the resonance and dynamics of the order-parameter amplitude in Eq.\,(\ref{eq:HiggsDyn_ConvSC})  are consistent with the long-time limit  reported in Refs.\,\cite{PhysRevB.101.184519,PhysRevB.92.064508}.

\begin{figure}
    \centering
    \includegraphics[width=\linewidth]{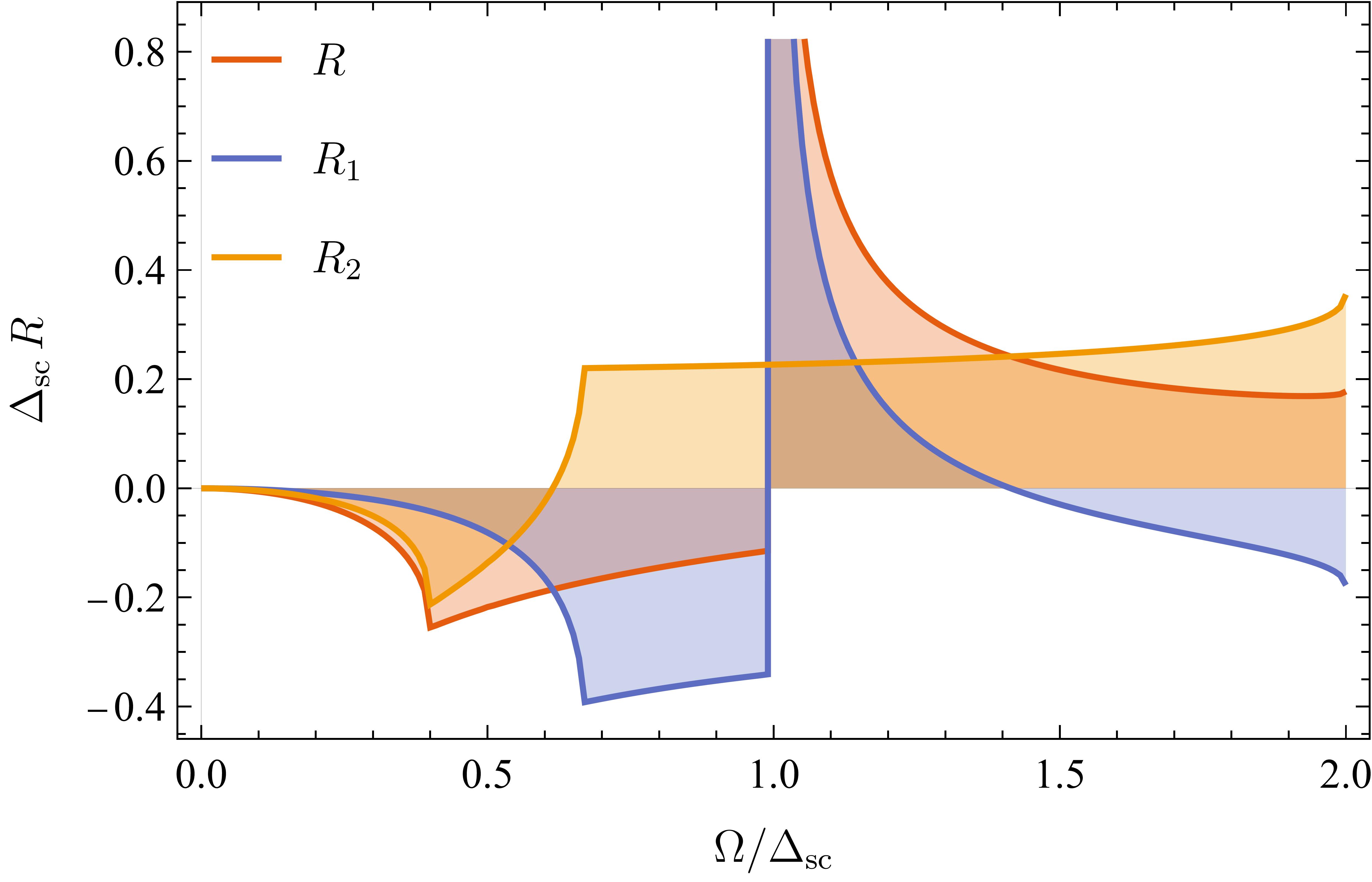}
    \caption{order-parameter amplitude $R$ as a function of the frequency of the drive $\Omega$, depicted by red curve. Blue and yellow curves show $R_{1}$  and $R_{2}$ as a function of $\Omega$, which correspond to  contributions   due to sidebands $n=\pm1$ and $n=0,\pm2$, respectively.   The amplitude $R$ becomes resonant at $\Omega=\Deltastatic$, as a result of the resonant behavior of the contributions due to $F_{\pm1,\mp1}$   $R_{1}$, see Eq.~\eqref{deltaD2}. Parameters: $\varepsilon_C=2000\Deltastatic$.  
    }
    \label{fig:AmplitudeSpectrum}
\end{figure}

To understand the dependence of $R(\Omega)$ on the driving frequency better, we plot the integrand of Eq.~\eqref{RR} as a function of $\xi_{\pmb{k}}$ and $\omega$ in Fig.~\ref{fig:FloquetBands}. In addition, therein we also plot the energies of the Floquet bands which follow from the zeros of $D_{n}$. We observe that $\tilde{R}_1+\tilde{R}_2$ acquires large values at the energies of the Floquet bands. We note that inside the integration boundaries, marked by gray dashed lines, the sidebands $n=\pm1$ give rise to a singularity of the integrand of the form $\sim1/(\Omega^2-\Deltastatic^2)^2$. Therefore, these Floquet bands give    the dominant contribution to the integral in Eq.~\eqref{RR}, which results in a resonance of $R(\Omega)$ at $\Omega=\Deltastatic$ as can be observed in Fig.~\ref{fig:AmplitudeSpectrum}, where we plot $R$ as a function of $\Omega$.
The resonant behavior of $R(\Omega)$ at $\Omega=\Deltastatic$ is directly reflected in the resonant behavior of $\Delta(t)$ in Eq.~\eqref{eq:HiggsDyn_ConvSC}. It is worth noting that the individual contributions $R_{1}$ and $R_{2}$, associated to sidebands $n=\pm1$ and $n=0,\pm2$, respectively, have a distinct impact on the total profile of the order-parameter amplitude determined by $R$. While  only $R_{1}$ reveals the resonance at $\Omega=\Delta_{\rm sc}$, both $R_{1}$ and $R_{2}$ develop kinks for $\Omega<\Delta_{\rm sc}$ which might cancel out as revealed by yellow and blue curves in Fig.~\ref{fig:AmplitudeSpectrum}. We find that these kinks  occur at finite frequencies of the drive given by 
\begin{equation}
\label{kinkOmega}
    \frac{\Omega}{\Deltastatic}=\frac{2}{2n+1}\,,
\end{equation} 
where $n$ is the leading sideband contribution. As already noted, incorporating more sidebands compensates the kinks and smooths out the overall features of $R$. Nevertheless,  the resonant profile at $\Omega=\Delta_{\rm sc}$ in the order-parameter amplitude $R$ remains strong, a signature associated to the Higgs resonance  which oscillates with $2\Omega$ as shown by Eq.~\eqref{eq:HiggsDyn_ConvSC}. Therefore, $R$ captures the Higgs dynamics. We note that such a resonant behavior of the order-parameter has been derived previously within the Anderson's pseudospin formalism~\cite{PhysRevB.92.064508}. Here, we have recovered such behavior purely by means of Floquet description including only a few Floquet bands.

\subsection{Nonequilibrium self-interaction}
\label{sec:NESI}
\begin{figure}
    \centering
    \includegraphics[width=\linewidth]{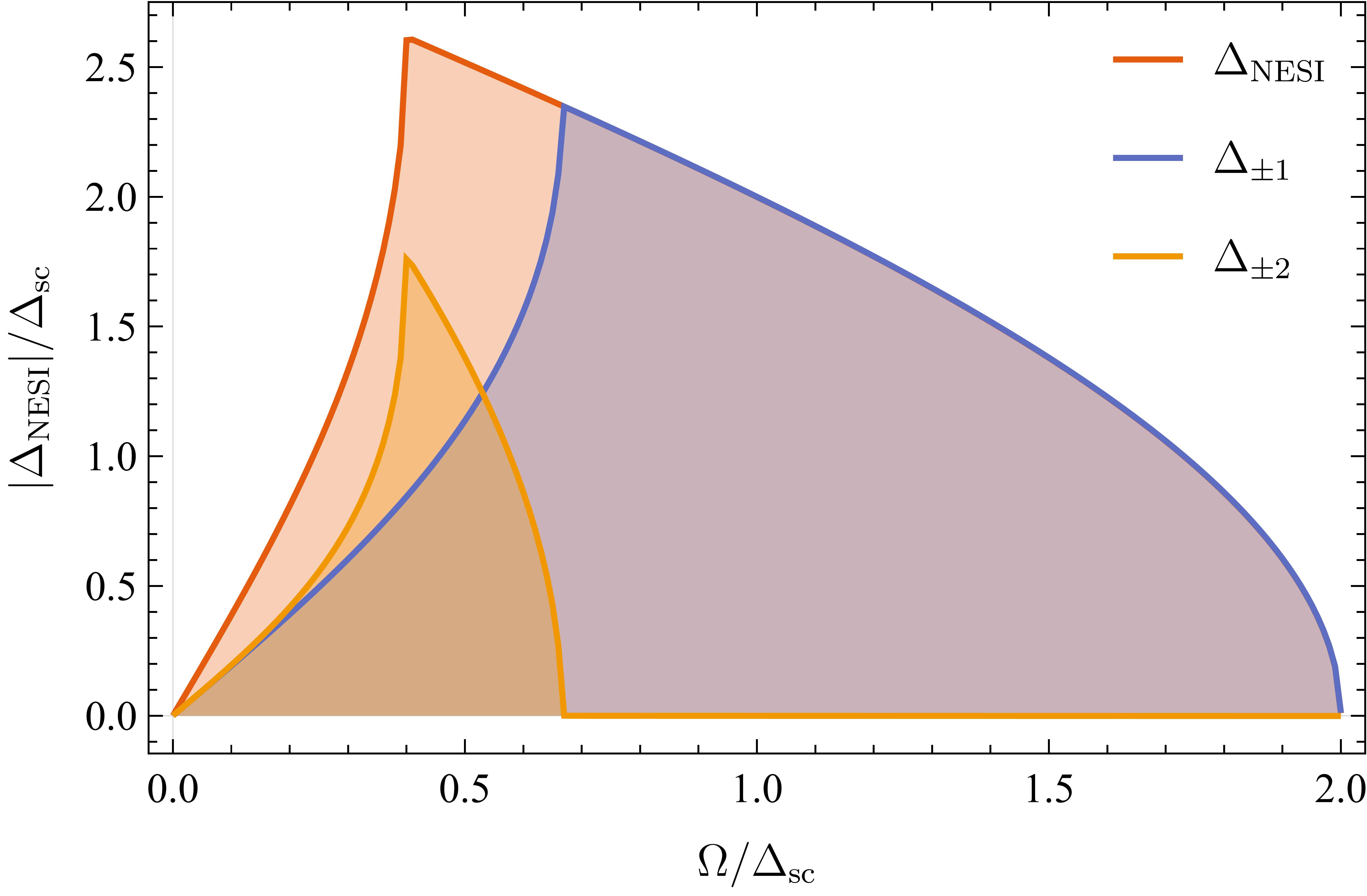}
    \caption{order-parameter of the NESI state as a function of the frequency of the drive $\Omega$, depicted by red curve. Blue and yellow curves represent the contributions to the NESI order-parameter due to nearest ($n=\pm1$) and next nearest ($n=\pm2$) sidebands. Beyond $\Omega=2\Delta_{\rm sc}$, the sideband spacing becomes large enough and the contribution to the NESI state is negligible. Parameters:   $\varepsilon_C=2000\Deltastatic$.}
    \label{fig:NESSSI}
\end{figure}
After we have analyzed the Higgs dynamics within Floquet theory, we now derive the NESI of the order-parameter following Eq.~\eqref{eq:NESI}.  Taking into account all Floquet modes, we arrive at 
\begin{equation}
    \label{NESIswave}
        \begin{split} \Delta_\text{NESI}=&\Delta_{0}(\Omega)-\Deltastatic\\
        =&i\lambda\int^{\varepsilon_C}_{-\varepsilon_C}\mathrm{d}\xi_{\pmb{k}}\int_{-\Omega/2}^{\Omega/2}\mathrm{d}\omega\;  \sum_{n\neq0}\frac{\Deltastatic}{D_n}+O(A_0^4)\,.
        \end{split}
    \end{equation}
Here, $D_{n}$ is given below Eq.~\eqref{FloquetpairF}, $\Deltastatic$ the static order-parameter, and $O(A_0^4)$ involve fourth- and higher-order corrections in the amplitude of the driving field $A_{0}$ given in Appendix~\ref{AppendixB}.  We note that the contribution of the central sideband with $n=0$ is given by
\begin{equation}
	\Deltastatic=\,i\lambda\int^{\varepsilon_C}_{-\varepsilon_C}\mathrm{d}\xi_{\pmb{k}}\int_{-\Omega/2}^{\Omega/2}\mathrm{d}\omega\;  \frac{\Deltastatic}{D_0},
\end{equation}
according to the BCS self-consistency equation and implies that the contribution of $n=0$ drops out of Eq.~\eqref{NESIswave}. 

In Fig.~\ref{fig:NESSSI}, we plot $\Delta_\text{NESI}$ as a function of the driving frequency $\Omega$. We also depict the individual contributions of the sidebands $n=\pm1$ and $\pm2$ (see the blue and yellow curves in Fig.~\ref{fig:NESSSI}). At $\Omega=0$, the NESI order-parameter vanishes because $\Delta_{0}(\Omega)=\Deltastatic$, in agreement with the expected behavior without a drive. As the driving frequency increases, $\Delta_\text{NESI}$ exhibits a growth and acquires a maximum (or kink) at drive frequencies below the static order-parameter, namely, $\Omega<\Deltastatic$. We have verified that the kink results from sideband contributions at driving frequencies where  the corresponding sideband touches the edge of the  integration bounds (see Fig.~\ref{fig:NESSSI}). In this regard, we note that each sideband produces a kink (maximum) such that the largest amplitude is associated to the sideband $n=\pm1$. The next sideband ($n=\pm2$) contribution in $\Delta_\text{NESI}$ develops a smaller maximum  at a  lower $\Omega$,  as shown by yellow curve in Fig.\,\ref{fig:NESSSI}. We have verified that higher sidebands ($n>\pm2$) form kinks at much lower amplitudes and at lower $\Omega$ which implies that the addition of many sidebands gives rise to a kink in $\Delta_\text{NESI}$ occurring at $\Omega$ given by Eq.\,(\ref{kinkOmega})

For larger driving frequencies, the NESI order-parameter decreases and vanishes for $\Omega\geq2\Deltastatic$, as a result of the central sideband  ($n=0$) touching the integration bounds, which then suppresses the  contributions from the remaining sidebands ($n>\pm1$). As higher Floquet band contributions are suppressed, the resulting steady-order-parameter is given by the static value $\Deltastatic$, as seen by Fig.\,\ref{fig:NESSSI}.  As $\Omega$ increases, the spacing between sidebands increases to a point $\Omega=2\Deltastatic$, where self-interactions between sidebands become negligible. As self-interactions, which are the key for the NESI state, vanish, it is expected that the order-parameter  of the superconducting system is simply given by the static one $\Deltastatic$, as is indeed observed in Fig.~\ref{fig:NESSSI}.  We have checked that by adding more sidebands the behavior of $\Delta_\text{NESI}$ at large frequencies remains unchanged.   The NESI order-parameter  discussed here thus provides evidence of a non-equilibrium superconducting phase with an order-parameter at a value that is distinct than $\Deltastatic$, which can be fully explained by Floquet picture.

\section{Conclusions}
\label{section5}
In conclusion, we presented a Floquet approach to study the Higgs dynamics in time-periodic superconductors, where the dynamics of the order-parameter is captured by Floquet pair amplitudes. 
We have shown that the Floquet description reduces the complexity of the time-dependent problem significantly. We illustrated our general theory with the example of a periodically driven conventional spin-singlet $s$-wave superconductor and showed that it correctly captures the Higgs mode as an order-parameter oscillation with twice the driving frequency which is resonant at the energy of the superconducting gap. In addition, we could show that the external driving gives rise to a renormalization of the static order-parameter component due to the coupling to higher Floquet bands, a signal that could, in principle, be measured. Our Floquet analysis of the Higgs dynamics can readily be extended to other superconducting systems such as spin-triplet superconductors as well as to the exploration of measurable observables in the presence of other competing effects as in Ref. \,\cite{PhysRevB.93.180507}. Given the ongoing efforts to control and manipulate the Higgs modes in superconductors, Floquet engineering provides a powerful method to understand order-parameter dynamics in driven superconducting materials.


\section{Acknowledgements}
We thank J. Zöllner and L. Litzba for insightful discussions. T. K. and B. S. acknowledge the financial support from the Deutsche Forschungsgemeinschaft (DFG, German Research Foundation), Project ID No. 278162697 – SFB 1242. J. C. acknowledges financial support from the Swedish Research Council  (Vetenskapsr\aa det Grant No.~2021-04121), and from Royal Swedish Academy of Sciences (Grant No.~PH2022-0003), and the Liljewalch travel grant.

 
\appendix

\section{Perturbative calculation of the Floquet Green's functions}
\label{AppendixA}
 We present further details of the Floquet Greens function and its components obtained within second-order perturbation theory. Focusing on the Floquet space spanned by $n=0,\pm1,\pm2$, the equation of motion gives rise to a Floquet Green's function given by
\begin{widetext}\begin{equation}\begin{split}
    \mathcal{G}_F(\pmb{k},\omega)&=\begin{pmatrix}
    \omega-H_\text{sc}-2\Omega&-\mathcal{U}^\ast_{1}&-\mathcal{U}^\ast_{2}&0&0\\
    -\mathcal{U}_{1}&\omega-H_\text{sc}-\Omega&-\mathcal{U}^\ast_{,1}&-\mathcal{U}^\ast_{2}&0\\
    -\mathcal{U}_{2}&-\mathcal{U}_{1}&\omega-H_\text{sc}&-\mathcal{U}^\ast_{1}&-\mathcal{U}^\ast_{2}\\
    0&-\mathcal{U}_{2}&-\mathcal{U}_{1}&\omega-H_\text{sc}+\Omega&-\mathcal{U}^\ast_{1}\\
    0&0&-\mathcal{U}_{2}&-\mathcal{U}_{1}&\omega-H_\text{sc}+2\Omega
    \end{pmatrix}^{-1}\\
    &=\begin{pmatrix}
    g_{-2,-2}^{-1}(\pmb{k},\omega)&-\mathcal{U}^\ast_{,1}&-\mathcal{U}^\ast_{,2}&0&0\\
    -\mathcal{U}_{,1}&g_{-1,-1}^{-1}(\pmb{k},\omega)&-\mathcal{U}^\ast_{,1}&-\mathcal{U}^\ast_{,2}&0\\
    -\mathcal{U}_{,2}&-\mathcal{U}_{,1}&g_{0,0}^{-1}(\pmb{k},\omega)&-\mathcal{U}^\ast_{,1}&-\mathcal{U}^\ast_{,2}\\
    0&-\mathcal{U}_{,2}&-\mathcal{U}_{,1}&g_{1,1}^{-1}(\pmb{k},\omega)&-\mathcal{U}^\ast_{,1}\\
    0&0&-\mathcal{U}_{,2}&-\mathcal{U}_{,1}&g_{2,2}^{-1}(\pmb{k},\omega)
    \end{pmatrix}^{-1}
\end{split}\label{eq:FloquetGreensFunction LinAndQuad}\end{equation}\end{widetext}
where the bare Green's functions $g_{n,n}(\pmb{k},\omega)$ on the diagonal can be written in Nambu space as
\begin{equation}
\begin{split}
    g_{0,0}(\pmb{k},\omega)&=(\omega-H_\text{sc})^{-1}\\
    &=\frac{1}{D_0}\begin{pmatrix}
    \omega+\xi_{\pmb{k}}&\Deltastatic\\
    \Deltastatic&\omega-\xi
    \end{pmatrix},\\
    g_{n,n}(\pmb{k},\omega)&=g_{0,0}(\pmb{k},\omega+n\Omega)\,,
    \label{eq:bare floquet green propagator}
\end{split}\end{equation}
with $D_0=\omega^2-\xi_{\pmb{k}}^2-\Deltastatic^2$ the determinant of $(\omega-H_\text{sc})$ and we have $H_{sc }$ given by Sec.~\ref{section4}. Due to the nature of the driving used in Sec.~\ref{section4}, only nearest and next-nearest-neighbor coupling between Floquet bands appear. The form of such coupling is explicitly shown in Eqs.~\eqref{eq:drive-potential} and~\eqref{eq:hopping-amplitudes} for a linearly polarized light drive.

\subsection{Components of the Floquet Green's function}
\label{sec:propagatorelements}
\begin{figure}
    \centering
    \includegraphics[width=\linewidth]{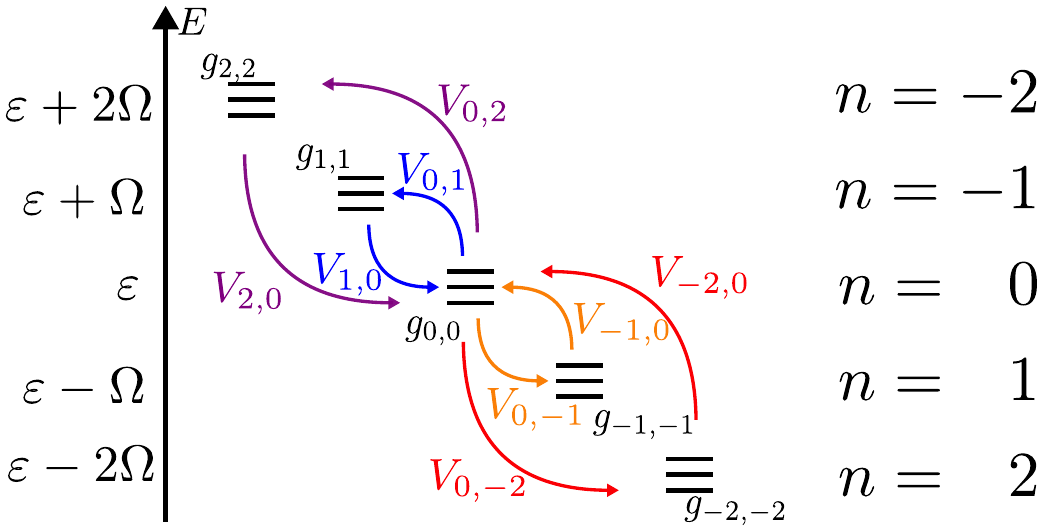}
    \caption{All available paths in Floquet sidebands for propagator $\mathcal{G}_{0,0}$ while hopping twice or less. This visual representation shows the physical interpretation of Eq.~\eqref{eq:G00}. The coupling $V_{r,s}$ are given by $V_{r,s}=-{U}_{s-r}$. Intra sideband interaction is represented by the bare propagator of the sidebands $g_{n,n}$. Here, $\varepsilon+n\Omega$ denotes the quasi energy of the $n$th Floquet sideband. Analog to this example, all available paths for the other propagator elements $\mathcal{G}_{n,m}$ need to be considered.}
    \label{fig:classdelta0}
\end{figure}

The elements of the Floquet Green's functions   can be determined by following the discussion presented in Sec.~\ref{subsection3e}, by using the Dyson's equation~\eqref{eq:dyson_GF} up to second order in the coupling between Floquet sidebands.    Projecting only on sidebands $n=0,\pm1,\pm2$, we  obtain the following elements:
 \begin{subequations}
    \begin{equation}\begin{split}
        \mathcal{G}_{0,0}=g_{0,0}+\sum_\pm (&g_{0,0}V_{0,\pm1}g_{\pm1,\pm1}V_{\pm1,0}g_{0,0}\\
        &+g_{0,0}V_{0,\pm2}g_{\pm2,\pm2}V_{\pm2,0}g_{0,0}),
        \label{eq:G00}        
    \end{split}
    \end{equation} 
    \begin{equation}
        \begin{split}   
            \mathcal{G}_{\pm1,\pm1}=g_{\pm1,\pm1}+&g_{\pm1,\pm1}V_{\pm1,\pm2}g_{\pm2,\pm2}V_{\pm2,\pm1}g_{\pm1,\pm1}\\+&g_{\pm1,\pm1}V_{\pm1,0}g_{0,0}V_{0,\pm1}g_{\pm1,\pm1}\\+&g_{\pm1,\pm1}V_{\pm1,\mp1}g_{\mp1,\mp1}V_{\mp1,\pm1}g_{\pm1,\pm1},
        \end{split}
    \end{equation}
    \begin{equation}
        \begin{split}
            \mathcal{G}_{\pm2,\pm2}=g_{\pm2,\pm2}+&g_{\pm2,\pm2}V_{\pm2,\pm1}g_{\pm1,\pm1}V_{\pm1,\pm2}g_{\pm2,\pm2}\\+&g_{\pm2,\pm2}V_{\pm2,0}g_{0,0}V_{0,\pm2}g_{\pm2,\pm2}.
        \end{split}
    \end{equation}
\end{subequations}
We note that each Floquet element above  involves intrasideband propagation $g_{n,n}$ and transitions between Floquet bands driven by $V_{m,n}$. Here, the couplings $V_{m,n}$ are obtained from  $V_{m,n}=-{U}_{n-m}$. The transitions between sidebands involve the absorption and emission of $n-m$ photon. In Fig.~\ref{fig:classdelta0}, we show an example of all the involved processes for 
  $\mathcal{G}_{0,0}$.

With the the diagonal elements of the Floquet pair amplitudes at hands, we can now write their off-diagonal components taking into account the difference $|m-n|$ which is useful for obtaining the dynamics of the order-parameter in Eq.~\eqref{eq:floquet_orderparameter}. Therefore, we obtain

\paragraph{$|m-n|=1$:}

\begin{subequations}
        \begin{equation}
        \begin{split} \mathcal{G}_{0,\pm1}=g_{0,0}V_{0,\pm1}g_{\pm1,\pm1}&+g_{0,0}V_{0,\pm2}g_{\pm2,\pm2}V_{\pm2,\pm1}g_{\pm1,\pm1}\\&+g_{0,0}V_{0,\mp1}g_{\mp1,\mp1}V_{\mp1,\pm1}g_{\pm1,\pm1},
    \end{split}\end{equation}
    \begin{equation}
        \begin{split}\mathcal{G}_{\pm1,0}=g_{\pm1,\pm1}V_{\pm1,0}g_{0,0}&+g_{\pm1,\pm1}V_{\pm1,\mp1}g_{\mp1,\mp1}V_{\mp1,0}g_{0,0}\\&+g_{\pm1,\pm1}V_{\pm1,\pm2}g_{\pm2,\pm2}V_{\pm2,0}g_{0,0},
  \end{split}  \end{equation}
    \begin{equation}
        \begin{split} \mathcal{G}_{\pm2,\pm1}=&g_{\pm2,\pm2}V_{\pm2,\pm1}g_{\pm1,\pm1}\\&+g_{\pm2,\pm2}V_{\pm2,0}g_{0,0}V_{0,\pm1}g_{\pm1,\pm1},
\end{split}    \end{equation}
    \begin{equation}
        \begin{split} \mathcal{G}_{\pm1,\pm2}=&g_{\pm1,\pm1}V_{\pm1,\pm2}g_{\pm2,\pm2}\\&+g_{\pm1,\pm1}V_{\pm1,0}g_{0,0}V_{0,\pm2}g_{\pm2,\pm2}.
    \end{split}\end{equation}
    \label{eq:l1propagator}
\end{subequations}

\paragraph{$|m-n|=2$}

\begin{subequations}
    \begin{equation}:
        \mathcal{G}_{0,\pm2}=g_{0,0}V_{0,\pm2}g_{\pm2,\pm2}+g_{0,0}V_{0,\pm1}g_{\pm1,\pm1}V_{\pm1,\pm2}g_{\pm2,\pm2},
    \end{equation}
    \begin{equation}
        \mathcal{G}_{\pm2,0}=g_{\pm2,\pm2}V_{\pm2,0}g_{0,0}+g_{\pm2,\pm2}V_{\pm2,\pm1}g_{\pm1,\pm1}V_{\pm1,0}g_{0,0},
    \end{equation}
    \begin{equation}
        \begin{split}
            \mathcal{G}_{\pm1,\mp1}=&g_{\pm1,\pm1}V_{\pm1,\mp1}g_{\mp1,\mp1}\\&+g_{\pm1,\pm1}V_{\pm1,0}g_{0,0}V_{0,\mp1}g_{\mp1,\mp1}.
    \end{split}\end{equation}
    \label{eq:l2propagator}
\end{subequations}

\paragraph{$|m-n|=3$:}
\begin{subequations}
    \begin{equation}
        \begin{split}
        \mathcal{G}_{\pm2,\mp1}=&g_{\pm2,\pm2}V_{\pm2,\pm1}g_{\pm1,\pm1}V_{\pm1,\mp1}g_{\mp1,\mp1}\\&+g_{\pm2,\pm2}V_{\pm2,0}g_{0,0}V_{0,\mp1}g_{\mp1,\mp1},
    \end{split}\end{equation}
    \begin{equation}\begin{split}
        \mathcal{G}_{\pm1,\mp2}=&g_{\pm1,\pm1}V_{\pm1,0}g_{0,0}V_{0,\mp2}g_{\mp2,\mp2}\\&
        +g_{\pm1,\pm1}V_{\pm1,\mp1}g_{\mp1,\mp1}V_{\mp1,\mp2}g_{\mp2,\mp2}\,.
            \end{split}
        \end{equation}
    \label{eq:l3propagator}
\end{subequations}

\paragraph{$|m-n|=4$:}

    \begin{equation}
        \mathcal{G}_{\pm2,\mp2}=g_{\pm2,\pm2}V_{\pm2,0}g_{0,0}V_{0,\mp2}g_{\mp2,\mp2}\,.
            \label{eq:l4propagator}
    \end{equation}
    
The components of the Floquet Green's function $\mathcal{G}_{n,m}$ obtained here are then used to find the Floquet pair amplitudes $F_{n,m}$, which are determined by the off-diagonal parts of $\mathcal{G}_{n,m}$.   This is what we carried out in Sec.~\ref{section4} when obtaining the dynamics of the order-parameter in a conventional spin-singlet $s$-wave time-periodic superconductor.

\subsection{order-parameter dynamics are real valued}
\label{sec:dDl proof}
In the paragraph below Eq.~\eqref{eq:floquet_orderparameter}, we discussed that the time-dependent order-parameter $\Delta(t)$ is real. Here we demonstrate this argument. 

We start with the definition of $\Delta_l(\Omega)$ given by Eq.~\eqref{eq:floquet_orderparameter3}:
\begin{equation}
        \Delta_{l}(\Omega)=iU\sum_{\pmb{k},m}\int_{-\Omega/2}^{\Omega/2}\mathrm{d}\omega\;  F_{m+l,m}(\pmb{k},\omega)\,.
\end{equation}
At this point,  we note that a sign change of $l\to-l$ is the equivalent to an index swap $m\leftrightarrow n$. Therefore, to show that the order-parameter is real, we need to prove that $F_{n,m}=F_{m,n}^*$. 

The Floquet pair amplitudes $F_{n,m}$ can be represented perturbatively in terms of Floquet Green's function $\mathcal{G}$ via Dyson's approach. Looking at Dyson's series Eq.~\eqref{eq:dyson_GF}, we have
\begin{equation}\begin{split}
    \mathcal{G}_{m,n}\approx&\langle m|g|n\rangle+\langle m|gVg|n\rangle+\langle m|gVgVg|n\rangle\\
                    =&\langle n|g|m\rangle^*+\langle n|gVg|m\rangle^*+\langle n|gVgVg|m\rangle^*\\
                    \approx&\,\mathcal{G}_{n,m}^*\,,
\end{split}\end{equation}
where $\mathcal{G}_{m,n}$ represent Floquet components. Now, because $F_{m,n}$ is a component of $\mathcal{G}_{m,n}$, the operation transfers to the order-parameter amplitudes $\Delta_l(\Omega)=\Delta_{-l}^*(-\Omega)$ which then shows that $\Delta_{0}(\Omega)$ is real.

Then, the time-dependent order-parameter can now be written as
\begin{equation}
\begin{split}
    \hat \Delta(t)=&\sum_{l}\Delta_l(\Omega) e^{il\Omega t}\\=&\Delta_{0}(\Omega)\\&+\sum_{l>0}2\big[\Real\{\Delta_l(\Omega)\}\cos{(l\Omega t)}\\
    &-\Imag\{\Delta_l(\Omega)\}\sin{(l\Omega t)}\big]\,,
\end{split}
\end{equation}
which implies that $\Delta(t)$ is a  real function. {This property is pointed out in Sec.\,\ref{section3} below Eq.\,(\ref{eq:floquet_orderparameter3}).}

\section{Higher order corrections to the NESI order-parameter}
\label{AppendixB}
In Sec.~\ref{sec:NESI} {we obtained the NESI order-parameter. We noted that it  includes intrasideband contributions as well as   terms containing fourth- and higher-order corrections in the amplitude of the driving field $A_{0}$. For completeness, here we write these corrections, which we obtain to be given by}
    \begin{equation}
        \begin{split}
            O(A_0^4)&=i\lambda\int^{\varepsilon_C}_{-\varepsilon_C}\mathrm{d}\xi_{\pmb{k}}\int_{-\Omega/2}^{\Omega/2}\mathrm{d}\omega\;\Deltastatic|\mathcal{U}_2|^2\\
            &\times\bigg[\frac{2(4\Omega^2+D_0)(4\xi_{\pmb{k}}^2+D_0)-8(2\omega\Omega)^2}{D_0^2D_{-2}D_{2}}\\
            &+\sum_\pm\bigg(\frac{D_{\pm1}+4\xi_{\pmb{k}}^2-4\Omega^2}{D_{\pm1}D_{\mp1}^2}+\frac{D_{0}+4\xi_{\pmb{k}}^2-4\Omega^2}{D_{0}D_{\pm2}^2}\bigg)\bigg]\,.            
        \end{split}
    \end{equation}
{It is straightforward to see that these corrections are proportional to $|\mathcal{U}_2|^2$. Then, by using the expressions for $\mathcal{U}_{2}$ from Eq.\,(\ref{eq:hopping-amplitudes}), we see that $|\mathcal{U}_2|^2\propto A_0^4$, which, for weak driving fields with small $A_{0}$, is insignificantly small when compared to the intra-sideband self-interaction in Eq.\,(\ref{NESIswave}).}


\bibliography{biblio}

\end{document}